\documentclass{article}
\usepackage{amsfonts,amssymb,xspace,pictex}

\newlength{\dinwidth}                                                          
\newlength{\dinmargin}                                                         
\makeatletter
\newif\if@fewtab\@fewtabtrue
\@addtoreset{equation}{section}
\def\cases#1{\left\{\,\vcenter{\normalbaselines\m@th
    \ialign{$\displaystyle{##}\hfil$&\quad##\hfil\crcr#1\crcr}}\right.}
\def\casetable#1{\left\{\,\vcenter{\normalbaselines\m@th
    \ialign{$\hfil\displaystyle{##}$&\quad##\hfil\crcr#1\crcr}}\right.}
\def\sideset#1#2#3{%
  \setbox\z@\hbox{$\displaystyle{\vphantom{#3}}#1{#3}\m@th$}%
  \setbox\tw@\hbox{$\displaystyle{#3}#2\m@th$}%
  \hskip\wd\z@\hskip-\wd\tw@\mathop{\hskip\wd\tw@\hskip-\wd\z@
  {\vphantom{#3}}#1\!{#3}#2}}
\makeatother

\newcommand{\Qn}{\mathbb{Q}}
\newcommand{\Zn}{\mathbb{Z}}
\newcommand{\Nn}{\mathbb{N}}

\newtheorem{Thm}{Theorem}

\newtheorem{Lem}{Lemma}

\newcommand{\lb}[1]{\label{#1}}
\newcommand{\Eq}[1]{(\ref{#1})}
\newcommand{\ct}[1]{\cite{#1}}

\renewcommand{\[}{\begin{eqnarray}}
\newcommand{\nn}{\nonumber}
\newcommand{\non}{\nonumber \\ }
\renewcommand{\]}{\end{eqnarray}}
\newcommand{\hh}[1]{\hspace*{#1}}

\newcommand{\een}{\end{enumerate}}
\newcommand{\ben}{\begin{enumerate}}

\newcommand{\FOR}[1]{\quad\mbox{for #1}}
\newcommand{\noi}{\noindent}
\newcommand{\ga}{\alpha}

\newcommand{\gd}{\delta}
\newcommand{\gep}{\epsilon}

\newcommand{\gl}{\lambda}

\newcommand{\gr}{\rho}

\newcommand{\gt}{\theta}
\newcommand{\gx}{\xi}

\newcommand{\gz}{\zeta}

\newcommand{\gL}{\Lambda}

\newcommand{\cl}{\ensuremath{\ell}\xspace}
\newcommand{\fg}{\ensuremath{\mathfrak{g}}\xspace}
\newcommand{\fh}{\mathfrak{h}}

\newcommand{\fp}{\ensuremath{\mathfrak{p}}\xspace}

\newcommand{\fW}{\mathfrak{W}}

\newcommand{\sT}{\mathcal{T}}
\newcommand{\sP}{\mathcal{P}}
\newcommand{\sX}{\mathcal{X}}
\newcommand{\sY}{\mathcal{Y}}
\newcommand{\va}{\mathbf{a}}
\newcommand{\vga}{\mbox{\boldmath$\ga$}}

\newcommand{\vgd}{\mbox{\boldmath$\gd$}\xspace}
\newcommand{\ve}{\mathbf{e}}
\newcommand{\vk}{\mathbf{k}}

\newcommand{\vgL}{\mbox{\boldmath$\gL$}\xspace}

\newcommand{\vp}{\mathbf{p}}
\newcommand{\vP}{\mathbf{P}}

\newcommand{\vX}{\mathbf{X}}
\newcommand{\vr}{\mathbf{r}}
\newcommand{\vs}{\mathbf{s}}

\newcommand{\vgt}{\mbox{\boldmath$\gt$}}

\newcommand{\vro}{\mbox{\boldmath$\gr$}}
\newcommand{\vgx}{\mbox{\boldmath$\gx$}}

\newcommand{\da}[1]{{ (\vgd\X\vga_{#1}) }}
\newfont\sgb{cmmib7} 

\newcommand{\svd}{{\hbox{\sgb \symbol{"0E}}}}

\newcommand{\Oint}{\oint\limits}

\newcommand{\Res}[2][]{\Oint_{#1}\!\frac{d#2}{2\pi i}\,}

\newcommand{\Vir}{\mathrm{Vir}}

\newcommand{\mod}{\,\mathrm{mod}\,}

\newcommand{\frc}[2]{{\textstyle \frac{#1}{#2}}}

\newcommand{\8}{\ensuremath{E_8}}
\newcommand{\9}{\ensuremath{E_9}}
\newcommand{\0}{\ensuremath{E_{10}}}

\renewcommand{\|}{\,|\,}
\renewcommand{\.}{\cdot}
\newcommand{\X}{\!\cdot\!}
\newcommand{\XO}{\otimes}

\newcommand{\1}{\mathbf{1}}

\newcommand{\ket}[1]{\ensuremath{|#1\rangle}}

\newcommand{\ord}[1]{\mbox{\large\bf:} #1 \mbox{\large\bf:}}
\newcommand{\xord}[1]{{}_\times^\times #1 {}_\times^\times}
\newcommand{\II}{I\hspace{-.2em}I_{9,1}}
\newcommand{\III}{I\hspace{-.2em}I_{25,1}}

\newcommand{\hv}{\ensuremath{h^\vee}\xspace}
\newcommand{\bfg}{\ensuremath{\bar{\fg}}\xspace}
\newcommand{\bfh}{\ensuremath{\bar{\fh}}\xspace}
\newcommand{\bvL}{\ensuremath{\bar{\vgL}}\xspace}
\newcommand{\bva}{\ensuremath{\bar{\va}}\xspace}
\newcommand{\bvr}{\ensuremath{\bar{\vro}}\xspace}
\newcommand{\bD}{\ensuremath{\bar{\Delta}}\xspace}
\newcommand{\hfg}{\ensuremath{\hat{\fg}}\xspace}
\newcommand{\K}[1][\cl]{\ensuremath{\sideset{^{\scriptscriptstyle[#1]}}%
                                             {_0}{K}}}
\newcommand{\vkl}{\ensuremath{\vk_\cl}\xspace}
\newcommand{\Lo}[1][\cl]{\ensuremath{\sideset{^{\scriptscriptstyle[#1]}}%
                                             {_0}{d}}}
\newcommand{\AI}[3][\cl]{\ensuremath{\sideset{^{\scriptscriptstyle[#1]}}%
                                             {_{#3}^{#2}}{\!A}}}
\newcommand{\Ai}[2][\cl]{\ensuremath{\sideset{^{\scriptscriptstyle[#1]}}%
                                             {_{#2}^i}{\!A}}}
\newcommand{\Aj}[2][\cl]{\ensuremath{\sideset{^{\scriptscriptstyle[#1]}}%
                                             {_{#2}^j}{\!A}}}
\newcommand{\AL}[2][\cl]{\ensuremath{\sideset{^{\scriptscriptstyle[#1]}}%
                                             {_{#2}^-}{\!A}}}
\newcommand{\Ar}[2][\cl]{\ensuremath{\sideset{^{\scriptscriptstyle[#1]}}%
                                             {_{#2}^\vr}{\!A}}}
\newcommand{\Er}[2][\cl]{\ensuremath{\sideset{^{\scriptscriptstyle[#1]}}%
                                             {_{#2}^\vr}{E}}}
\newcommand{\Es}[2][\cl]{\ensuremath{\sideset{^{\scriptscriptstyle[#1]}}%
                                             {_{#2}^\vs}{E}}}
\newcommand{\Erb}[2][\cl]{\ensuremath{\sideset{^{\scriptscriptstyle[#1]}}%
                                             {_{#2}^{\vr+\vs}}{E}}}
\newcommand{\Ema}[2][\cl]{\ensuremath{\sideset{^{\scriptscriptstyle[#1]}}%
                                             {_{#2}^{-\vr}}{E}}}
\newcommand{\XI}[1][{}]{\ensuremath{\sideset{^{\scriptscriptstyle[\cl]}}%
                                             {^{#1}}{\sX}}}
\newcommand{\PI}[1][{}]{\ensuremath{\sideset{^{\scriptscriptstyle[\cl]}}%
                                             {^{#1}}{\sP}}}
\newcommand{\PL}[2][\cl]{\ensuremath{\sideset{^{\scriptscriptstyle[#1]}}%
                                             {^{#2}}{\sP}}}
\newcommand{\YI}[2][{}]{\ensuremath{\sideset{^{\scriptscriptstyle[\cl]}}%
                                             {^{#1}_{#2}}{\sY}}}
\newcommand{\sL}[2][\cl]{\ensuremath{\mathcal{L}^{[#1]}_{#2}}}
\newcommand{\sK}[2][\cl]{\ensuremath{\mathcal{K}^{[#1]}_{#2}}}
\newcommand{\tsL}[2][\cl]{\ensuremath{\tilde{\mathcal{L}}^{[#1]}_{#2}}}

\newcommand{\LV}[2]{\ensuremath{L^{(#1)}_{#2}}}
\newcommand{\cs}[1]{\langle#1\rangle}
\begin{document}
\arraycolsep3pt

\thispagestyle{empty}
\renewcommand{\thefootnote}{\fnsymbol{footnote}}
\begin{flushright} hep-th/9604155 \\
                   DESY 96-072    \end{flushright}
\vspace*{2cm}
\begin{center}
{\LARGE \sc The Sugawara generators at arbitrary level%
            \footnote[1]{submitted to \emph{Commun.\ Math.\ Phys.}}}\\
 \vspace*{1cm}
       {\sl R.~W. Gebert\footnote[2]{Supported by Deutsche
    Forschungsgemeinschaft under Contract No.\ \emph{DFG Ni 290/3-1}}
            \qquad K. Koepsell \qquad H. Nicolai}\\
 \vspace*{6mm}
     II. Institut f\"ur Theoretische Physik, Universit\"at Hamburg\\
     Luruper Chaussee 149, D-22761 Hamburg, Germany\\
 \vspace*{6mm}
 \date\\
\vspace*{1cm}
\begin{minipage}{11cm}\footnotesize
  {\bf Abstract:} We construct an explicit representation of the
  Sugawara generators for arbitrary level in terms of the homogeneous
  Heisenberg subalgebra, which generalizes the well-known expression
  at level 1. This is achieved by employing a physical vertex operator
  realization of the affine algebra at arbitrary level, in contrast to
  the Frenkel--Kac--Segal construction which uses unphysical
  oscillators and is restricted to level 1. At higher level, the new
  operators are transcendental functions of DDF ``oscillators'' unlike
  the quadratic expressions for the level-1 generators. An essential
  new feature of our construction is the appearance, beyond level 1,
  of new types of poles in the operator product expansions in addition
  to the ones at coincident points, which entail (controllable)
  non-localities in our formulas. We demonstrate the utility of the
  new formalism by explicitly working out some higher-level examples.
  Our results have important implications for the problem of
  constructing explicit representations for higher-level root spaces
  of hyperbolic Kac--Moody algebras, and $\0$ in particular.

\end{minipage}
\end{center}
\renewcommand{\thefootnote}{\arabic{footnote}}
\setcounter{footnote}{0}
\newpage

\section{Introduction} \lb{sec:Int}
The Sugawara construction \ct{Suga68} and the GKO construction
\ct{GoKeOl85} have both come to play a prominent role in 
string theory and in the theory of Kac--Moody algebras
(see e.g.\ \ct{Kac90,MooPia95} for the general theory). As is well
known, the Sugawara construction extends a representation
of an affine Lie algebra \fg to that of its semidirect product with
the Virasoro algebra $\Vir^\fg$.  The GKO construction, in turn, is
based on the Sugawara construction: given an affine subalgebra
$\fp\subset\fg$, there always exists another Virasoro algebra
corresponding to the difference of Virasoro operators associated with
\fg and \fp, respectively, such that the resulting coset Virasoro
algebra $\Vir^{\fg,\fp}$ commutes with the affine subalgebra \fp. It is
for this reason that the GKO construction, which was originally
developed for the explicit description of $c<1$ Virasoro
modules \ct{GoKeOl85}, has acquired great importance in
the representation theory of affine algebras \ct{KacWak89}. More
specifically, every highest weight representation $L(\gL)$ of \fg can be
decomposed w.r.t.\ the direct sum $\fp\oplus\Vir^{\fg,\fp}$ as
follows:
\[ L(\gL)=\bigoplus_{\gl\in P_+^{\fg,\fp}}L^\fp(\gl)\XO U(\gL,\gl). 
              \lb{prod-rep1} \]
The relevant specialization of this formula for tensor
products of \fg modules is obtained by taking \fp to be the diagonal
subalgebra of $\fg \otimes {\bf 1} \oplus {\bf 1} \otimes \fg$.
In this case \Eq{prod-rep1} becomes
\[ L(\gL)\XO L(\gL') = \bigoplus_{j}
   L(\gL_j)\XO V(c_j,h_j), \lb{prod-rep2}  \]
where $V(c_j,h_j)$ is the Virasoro module with central charge $c_j$
and highest weight $h_j$. Unfortunately, it is not so easy in general
to compute the products \Eq{prod-rep2} in practice. For simple
examples, where $c_j<1$ and unitarity restricts $h_j$ to a finite set
of allowed values, one can work out the product explicitly. However,
at higher level the right-hand side of \Eq{prod-rep2} will contain
many terms; furthermore, these will in general correspond to Virasoro
Verma modules with central charge $c_j>1$ where the values of $h_j$
are unrestricted (apart from $h_j \ge0$). In order to better
understand these representations it is thus desirable to find an
explicit and manageable representation for the coset generators. A
necessary prerequisite for this is the construction of an explicit
representation for the higher-level Sugawara operators themselves (the
explicit representation at level~1 is well known, see formula
\Eq{lev1-equiv} below). This is the problem which we address and solve
in the present paper.

Before going into the details we would like to briefly explain the new
features of our construction. The famous FKS vertex operator
realization \ct{FreKac80,Sega81} of nontwisted affine Lie algebras
corresponds to a spatially compactified bosonic string whose momentum
lattice is taken to be the (Euclidean) root lattice of a
finite-dimensional simple Lie algebra of $ADE$ type. The Laurent
coefficients (``modes'') of the tachyon vertex operators together with
the string oscillators then constitute a basis of the affine algebra.
This basis, however, is not physical in the sense of string theory since 
except for the zero mode these mode operators do not commute
with the Virasoro constraints. Furthermore, the FKS construction has
the drawback of being restricted to affine algebras at level~1. In
\ct{Fren85,GodOli85}, it was noticed that if the momentum lattice of
the string is enlarged by a two-dimensional Minkowski lattice then the
zero mode operators by themselves already lead to a basis of the affine
algebra which agrees with the FKS realization when the operator-valued
expression $e^{in\svd\.\vX(z)}$ is formally replaced by $z^n$. Our
starting point was the observation that apart from being manifestly
physical, this construction is applicable to affine Lie algebras at
arbitrary level and thus more general than the FKS construction (this
fact is apparently not widely known).  

In the usual vertex operator formalism the computation of Lie algebra
commutators is reduced to the evaluation of the singular terms in the
expansion of certain operator products at coincident points. By
contrast, we here find that, beyond level~1, additional poles appear
at the non-coincident points $z=w_p:= \zeta^p w$ in the expansion of
the product of two conformal operators supported at $z$ and $w$, where
$\zeta$ is a primitive $\cl$-th root of unity.  A second unusual
feature is that we are led to introduce a new ``transversal
coordinate'' field, which has the form of the old Fubini-Veneziano
field, but for which the usual string oscillators are replaced by the
level-$\cl$ DDF operators, see Eq.\ \Eq{DDF-coo} below.  The final
expression for the general level-$\cl$ Sugawara operators (cf.\ \Eq{Sug-4}
below) involves an exponential dependence on this new field; since the
DDF operators themselves are defined by exponentials of the string
oscillators, our construction may thus be termed ``doubly
transcendental''. Moreover it is {\em non-local} in the sense that the
integrand in \Eq{Sug-4} below depends both on $w$ and $w_p$.  This
(controllable) non-locality accounts for a number of complications,
such as the fact that the level-$\cl$ Sugawara operators contain
products of the DDF operators of arbitrary order (depending on the
state on which they act) whereas the level-1 Sugawara operators
are always quadratic. 

The present work is mainly motivated by and continues our previous
investigation of hyperbolic Kac--Moody algebras corresponding to the
canonical extensions of affine algebras by an over-extended root
$\vr_{-1}$ \ct{GebNic95,GebNic94}. There an attempt was made to
understand the structure of such algebras, and in particular the
maximally extended algebra \0, via a novel realization in terms of DDF
states which enabled us to give a simple and explicit representation
for a non-trivial level-2 root space of $E_{10}$ corresponding to a
75-fold multiple commutator of the Chevalley--Serre generators for the
first time (meanwhile, further examples have been worked out). These
results explicitly demonstrate the occurrence of {\it longitudinal}
states for levels $|\cl|\geq 2$ and the simultaneous decoupling of
certain transversal states, whereas the level~$\pm 1$ sectors can be
simply realized as the set of purely transversal states
\ct{GebNic95,Fren85} (the level-0 sector is just the affine
subalgebra). Let us recall that the higher-level elements of the
algebra can be recursively defined as multiple commutators of level-1
elements. In principle one should thus be able to understand them from
a representation theoretic point of view by analyzing multiple
products of level-1 representations. However, a first difficulty here
is that one must discard all those multiple commutators (and hence the
corresponding affine representations) which contain the Serre
relations somewhere inside. This difficulty is invisible in the string
vertex algebra realization \ct{Borc86}, which takes automatic care of
the Serre relations (since there are no physical string states below
the tachyon), but the tribute to this convenience is the phenomenon of
``missing'' (or ``decoupled'') states, i.e., physical string states
that can {\em not} be reached by multiple commutation of the
Chevalley--Serre generators \ct{GebNic95}. The second difficulty ---
already alluded to above --- is that there is no general method for
efficiently computing the relevant products of representations in
practice. So we see again that we must gain a better understanding of
the higher-level Sugawara generators. Although the challenge of
finding explicit formulas for the coset Virasoro generators remains,
we believe that the present results bring us one step closer towards
the ambitious goal of finding a concrete realization of hyperbolic
Kac--Moody algebras. With this future application in mind, we will
give explicit examples of the new formula only for some special
representations of \9 which arise in the analysis of the hyperbolic
algebra $\0$.

Apart from the new structural insights afforded by the new formula,
our results illustrate to what degree the higher-level sectors of
hyperbolic Kac--Moody algebras are more complicated than the low-level
sectors. This is acutely evident from the increasing ``anisotropy" of
the higher-level root spaces, which was already observed in
\ct{GebNic95}, and which is now (partially) explained by the symmetry
breaking of the full (affine) Weyl group down to a finite (and
generically trivial) subgroup called ``little Weyl group" in
\ct{GebNic95}. From a more technical perspective this phenomenon is
due to the appearance of certain weighted sums of tensor products of
roots (see \Eq{wsum-4}) which have not yet appeared in the literature
to the best of our knowledge.  We would like to emphasize that these
and other special features of hyperbolic Kac--Moody algebras can {\em
  not} be explained by string compactification alone.  In other words,
such algebras reveal an enigma beyond the string vertex operator
construction. By contrast, many of the generalized Kac--Moody
(super)algebras which have recently received attention (see e.g.\ 
\ct{Borc95,GriNik95,HarMoo95}) can presumably be realized as
untruncated, hence {\it bona fide}, algebras of physical vertex
operators of some compactified string. For instance, the so-called
fake monster Lie algebra is just the algebra of \emph{all} transversal
physical states of the bosonic string compactified on $\III$; its root
spaces are therefore perfectly ``isotropic" and the associated root
multiplicities are simply given by the number of physical string
states.

\section{Preliminaries} \lb{sec:Set}
We consider a nontwisted affine Lie algebra \fg with underlying simple 
finite-dimensional Lie algebra \bfg of rank $d-2$ ($d\ge3$). The
associated hyperbolic Kac--Moody algebra \hfg of rank $d$ is obtained
by adjoining to the affine Dynkin diagram another node which is
related to the over-extended simple root $\vr_{-1}$. The extended
(Minkowskian) affine root lattice is defined as
$\hat{Q}:=Q\oplus\Zn\vgL_0$ (viewed as the even sublattice of the
affine weight lattice), where $Q$ denotes the affine root lattice and
$\vgL_0:=\vr_{-1}+\vgd$ is a null vector conjugate to the affine null
root $\vgd$. Clearly, $\hat{Q}$ is just the hyperbolic root lattice.
For any element \vgL of $\hat{Q}$, the level~\cl is defined by
\[ \cl:=-\vgL\X\vgd. \]
Now suppose that $\vgL\in\hat{Q}$ is a root of \hfg of nonzero
level. The DDF decomposition of \vgL \ct{GebNic95} is defined by 
\[ \vgL = \va - n \vkl,  \lb{DDF-dec}  \]
where we have put
\[ \vkl := -\frac1\cl\vgd \quad\in Q_\Qn:=\Qn\XO_{\Zn}Q, \lb{k-def} \]
and where the vector $\va$ is uniquely determined by requiring
$\va^2=2$, i.e., $n:=1-\frc12\vgL^2$. Note that $n$ is always a
non-negative integer since $\vgL^2\le2$ for any root. We will refer to
$\va$ as the `tachyonic level-\cl vector' and to $\ket\va$ as the
`tachyonic level-\cl state' associated to \vgL; beyond level~1, it
need not be an element of the lattice $\hat{Q}$ but only of its
rational extension.  Let us furthermore introduce the orthonormal
polarization vectors $\vgx_i \equiv \vgx_i (\va)$ satisfying
$\vgx_i\X\vgx_j=\gd_{ij}$ and $\vgx_i\X\vgd=\vgx_i\X\vgL=\vgx_i\X\va
=0$. They constitute a basis of the complex vector space $\bfh^*$ dual
to the Cartan subalgebra $\bfh$ of \bfg. Then we define the operators
\[ \Ai{m}(\va) 
    &:=& \Res{z}\vgx_i(\va)\X\vP(z)e^{im\vkl\.\vX(z)}\lb{CWbas-1} \\
   \Er{m} 
    &:=& \Res{z}\ord{e^{i(\vr+m\vkl)\.\vX(z)}}, \lb{CWbas-2} \]
for $m\in\Zn$, $1\le i\le d-2$, $\vr\in\bD$ (roots of \bfg). 
Here we have used the well-known Fubini--Veneziano coordinate 
and momentum fields, respectively,
\[ X^\mu(z) &:=& 
   q^\mu-ip^\mu\ln z+i\sum_{m\ne0}\frac1m\ga^\mu_mz^{-m}, 
   \lb{FuVe-coo} \\
   P^\mu(z) &:=& i\frac{d}{dz}X^\mu(z)
              = \sum_{m\in\Zn}\ga^\mu_mz^{-m-1}, \lb{FuVe-mom} \]
expressed in terms of the string oscillators $\ga^\mu_m$ ($m\in\Zn$,
$1\le\mu\le d$), 
\[ [\ga^\mu_m,\ga^\nu_n]=m\eta^{\mu\nu}\gd_{m+n,0}. \] 
The shift of any polarization vector $\vgx_i(\va)$ along the \vgd
direction leaves the associated DDF operator $\Ai{m}(\va)$ unchanged
for $m\ne0$, because the residue of a total derivative always vanishes.
Since the polarization vectors of two tachyonic level-\cl states
are related by
\[ \vgx_i(\va')=\vgx_i(\va)+\frac1\cl\big(\vgx_i(\va)\X\va'\big)\vgd,  \]
we are thus effectively dealing with a single set of 
DDF operators for $m\neq 0$,
\[ \Ai{m}\equiv\Ai{m}(\va)=\Ai{m}(\va'); \lb{a-cl} \] 
so we can suppress the labels $\va$, $\va'$ in the remainder, i.e.\
write $\Ai{m}\equiv\Ai{m}(\va)$. Let us stress, however, that the zero
mode operators do differ for different $\va$, viz.\ 
\[ \Ai{0}(\va)=\vgx_i(\va)\X\vp
    =\vgx_i(\va')\X\vp-\frac1\cl\big(\vgx_i(\va)\X\va'\big)\vgd\X\vp
    =\Ai{0}(\va')-\frac1\cl\big(\vgx_i(\va)\X\va'\big)\vgd\X\vp. \]
For definiteness, we choose the polarization vectors to be
$\vgx_i(\vgL_0)$ throughout this paper, where $\vgL_0$ denotes the
above fundamental dominant weight of level~1 with tachyonic level-1
vector $\va_0=\vgL_0-2\vgd\in\hat{Q}$. This vector $\va_0$ plays the
role of the simple root $\vr_{-1}$ occurring in the canonical
extension of $\fg$ to the hyperbolic Kac--Moody algebra $\hfg$ with
$\hat{Q}$ as root lattice. Last not least we have the obvious relation
\[ \Ai[1]{m}(\va) = \Ai{\cl m} (\va).  \lb{1-cl}  \]

The above operators obey the commutation relations
\[ [\Ai{m},\Aj{n}] &=& m\gd^{ij}\gd_{m+n,0}\K,
    \lb{CWcom-1} \\[.5ex]
   [\Ai{m},\Er{n}] &=& (\vgx_i\X\vr)\Er{m+n}, \lb{CWcom-2} \\[.5ex]
   [\Er{m},\Es{n}] &=& 
     \cases{ 0 & if $\vr\X\vs\ge0$, \cr
            \gep(\vr,\vs)\Erb[\cl]{m+n} & if $\vr\X\vs=-1$, \cr
            -\Ar{m+n}-m\gd_{m+n,0}\K & if
            $\vr\X\vs=-2$,} \lb{CWcom-3} \]
where $\K := \vkl\X\vp = -\frac1\cl\vgd\X\vp$ denotes the operator
realization of the central element of the affine algebra and 
\[ \Ar{m} &:=& \Res{z}\vr\X\vP(z)e^{im\vkl\.\vX(z)} 
                = \sum_{i}(\vgx_i\X\vr)\Ai{m}
   \qquad\forall\vr\in\bD. \nn \]
As usual, we have to extend the Cartan subalgebra by an exterior
derivative which we choose to be $\Lo := \cl(\vgL_0+\vgd)\X\vp$ for the
basic (level~1) fundamental weight $\vgL_0$ (note that
$(\vgL_0+\vgd)^2=0$).

The operators \K[1], \Lo[1], \Ai[1]{m}, \Er[1]{m} ($1\le
i\le d-2$, $\vr\in\bD$, $m\in\Zn$) establish a realization of the
affine Lie algebra \fg, the level being given by the eigenvalue of the
operator \K[1]. Note that in contrast to the FKS construction this vertex
operator realization works for \emph{arbitrary level} and is
\emph{physical} in the sense of string theory, i.e.
\[  [L_m,\K[1]]=[L_m,\Lo[1]]=[L_m,\Ai[1]{n}]=[L_m,\Er[1]{n}]=0 
    \qquad\forall m,n\in\Zn, 1\le i\le d-2, \vr\in\bD, \nn \] 
where the operators
\[ L_m:=\frc12\sum_{n\in\Zn}\ord{\vga_n\X\vga_{m-n}}. \]
satisfy the standard Virasoro algebra with central charge $c=d$.
There is yet another realization of the affine Lie algebra which is,
however, restricted to level~1. Namely, on states with eigenvalue \cl
for \K[1], the operators \K, \Lo, \Ai{m}, \Er{m} ($1\le i\le d-2$,
$\vr\in\bD$, $m\in\Zn$) form a level-1 realization of \fg which is
also physical. 

Since we are working with the so-called homogeneous vertex operator
construction we will refer to the algebra of operators \Ai{m} as the
\textbf{homogeneous Heisenberg subalgebra} of the affine algebra. The
crucial observation for our analysis is that these operators not only
occur as part of the affine algebra but also as part of the spectrum
generating algebra for the physical string states. In this context,
they are nothing but the well-known transversal DDF operators. 
A crucial new feature of our analysis is the appearance of the
level-\cl \textbf{transversal coordinate field}
\[ \XI[i](z) := 
   q^i-ip^i\ln z+i\sum_{m\ne0}\frac1m\Ai{m}z^{-m},
   \lb{DDF-coo} \]
and the level-\cl \textbf{transversal momentum field} 
\[ \PI[i](z) := iz\frac{d}{dz}\XI[i](z)
             = \sum_{m\in\Zn}\Ai{m}z^{-m}, \lb{DDF-mom} \]
respectively, neither of which has appeared in the literature so far. 
Evidently, these fields are transcendental expressions in terms of 
the standard oscillator basis. The momentum field \Eq{DDF-mom} is physical 
because it commutes with the Virasoro constraints term by term.
This is not quite true of \Eq{DDF-coo} due to the presence of 
the center of mass coordinate $q^i$ in it; however, in all
relevant expressions below we will be dealing with the fields
\[ \YI[i]{p}(z)  := \XI[i](z_p)-\XI[i](z),\qquad p=1,\ldots,\cl-1\,,
    \lb{DDF-diff} \]
where $z_p:=\gz^pz$ and $\gz$ denotes a primitive \cl-th root of
unity. These fields are physical since the zero mode $q^i$ drops out. 
      
The Sugawara generators built from the affine Cartan--Weyl basis
\Eq{CWcom-1}-\Eq{CWcom-3} are
\[ \sL{m} := \frac1{2(\cl+\hv)}\sum_{n\in\Zn}
            \bigg(\sum_{i=1}^{d-2}\xord{\Ai[1]{n}\Ai[1]{m-n}}
            +\sum_{\vr\in\bD}\xord{\Er[1]{n}\Ema[1]{m-n}}\bigg),
            \lb{Sug-def} \]
where \hv denotes the dual Coxeter number of \bfg.
Normal-ordering is defined by 
\[ \xord{\Ai[1]{m}\Aj[1]{n}} 
         &:=& \cases{ \Ai[1]{m}\Aj[1]{n} & for $m\le n$, \cr
                      \Aj[1]{n}\Ai[1]{m} & for $m>n$, } \lb{nopx-1} \\
    \xord{\Er[1]{m}\Es[1]{n}} 
         &:=& \cases{ \Er[1]{m}\Es[1]{n} & for $m\le n$, \cr 
                      \Es[1]{n}\Er[1]{m} & for $m>n$. } \lb{nopx-2} \] 
It is well known that the operators \sL{m}, $m\in\Zn$, form a
Virasoro algebra (see e.g.\ \ct{GodOli86} and references therein),
\[ [\sL{m},\sL{n}]=(m-n)\sL{m+n}
                   +\frac{c(\cl)}{12}(m^3-m)\gd_{m+n,0}\K,
   \lb{Sug-Vir} \]
with central charge
\[ c(\cl):=\frac{\cl\dim\bfg}{\cl+\hv}. \lb{Sug-cc} \]
These operators act as outer derivations on the affine algebra so that
one obtains a semidirect product $\Vir_{\sL{}}\ltimes\fg$:
\[ [\sL{m},\Ai[1]{n}]=-n\Ai[1]{m+n},\qquad 
   [\sL{m},\Er[1]{n}]=-n\Er[1]{m+n}. \lb{sem-prod} \]
By construction, the Sugawara generators are physical, viz.
\[ [\sL{m},L_n]=0\qquad\forall m,n\in\Zn. \]
Thus the above semidirect product is a symmetry of the physical string
spectrum, whereas in the FKS approach only the full Fock space carries
a (level-1) representation of the affine algebra. It should be
mentioned that in addition to the operators $\sL{m}$, there is another
 infinity of ``physical" Virasoro algebras (but with uniform central charge
$c=26-d$) generated by the longitudinal DDF operators $\AL{m}(\va)$, 
all of which commute with the Sugawara generators \Eq{Sug-def}.
However, we will not elaborate on this point here; for further 
information, the interested reader may consult \ct{GebNic95}.

\section{The main formula} \lb{sec:Cal}
Our aim is to rewrite the Sugawara generators \sL{m} in terms of the
homogeneous Heisenberg subalgebra spanned by the \Ai{m}'s. This will be
the generalization of the well-known result
\[ \sL[1]{m}=\frac12\sum_{n\in\Zn}
            \sum_{i=1}^{d-2}\xord{\Ai[1]{n}\Ai[1]{m-n}},
            \lb{lev1-equiv} \]
which is referred to in the literature as `the equivalence of the
Sugawara and the Virasoro construction at level~1'.\footnote{Although,
  strictly speaking, this statement has only been proven in the
  framework of the FKS construction.}
For this purpose, we wish to evaluate the operator products occurring
in the second part of the Sugawara generators. We start from the
well-known formulas
\[ \ord{e^{i(\vr+n\svd)\.\vX(z)}}\,
   \ord{e^{-i(\vr+(m+n)\svd)\.\vX(w)}} 
   &=& (z-w)^{-2}\ord{e^{i\vr\.[\vX(z)-\vX(w)]}}
         e^{-im\svd\.\vX(w)} 
         e^{in\svd\.[\vX(z)-\vX(w)]} \non
   \ord{e^{-i(\vr+(m-n)\svd)\.\vX(w)}}\,
   \ord{e^{i(\vr-n\svd)\.\vX(z)}}
   &=& (z-w)^{-2}\ord{e^{i\vr\.[\vX(z)-\vX(w)]}}
         e^{-im\svd\.\vX(w)} 
         e^{-in\svd\.[\vX(z)-\vX(w)]}, \nn \]
where the exponentials involving \vgd need not be normal ordered since
$\vgd^2=\vgd\X\vr=0$. Invoking the algebraic identity
\[ \sum_{n>0}q^{-n}=(1-q^{-1})^{-1}-1
                   =-(1-q)^{-1}
                   =-\sum_{n\ge0}q^n, \nn \]
we get
\[ \lefteqn{\sum_{\vr\in\bD}\sum_{n\in\Zn}
            \xord{\Er[1]{n}\Ema[1]{m-n}}} \hh{8mm} \non
   &=& \sum_{\vr\in\bD}\bigg[
        \Res[0]{w}\Res[|z|>|w|]{z}\sum_{n\ge0}
         \ord{e^{i(\vr+n\svd)\.\vX(z)}}\,
         \ord{e^{-i(\vr+(m+n)\svd)\.\vX(w)}} \non
   & & {}\hh{10mm}+\Res[0]{w}\Res[|z|<|w|]{z}\sum_{n>0}
         \ord{e^{-i(\vr+(m-n)\svd)\.\vX(w)}}\,
         \ord{e^{i(\vr-n\svd)\.\vX(z)}}\bigg] \non
   &=& \sum_{\vr\in\bD}\Res[0]{w}
       \sum_{w_p\in\{\mathrm{poles}\}}\Res[z=w_p]{z}
         (z-w)^{-2}\ord{e^{i\vr\.[\vX(z)-\vX(w)]}}
         e^{-im\svd\.\vX(w)} 
         \sum_{n\ge0}e^{in\svd\.[\vX(z)-\vX(w)]}, \lb{adad-1} \]
where the second sum runs over all poles of the integrand in the region
\[ C_w:=\lim_{\gep\to0}\{z\||w|-\gep\le|z|\le|w|+\gep\}
       =\{z\||z|=|w|\}, \nn \]
i.e., on a circle of radius $|w|$ in the $z$ plane. 
A crucial observation for our construction is that besides the
the obvious pole at $z=w$, there will be extra poles for level
$|\cl|\geq2$ in the operator-valued function
\[ Y(z,w):=\sum_{n\ge0}e^{in\svd\.[\vX(z)-\vX(w)]}. \nn \]
These are due to the replacement of $(w/z)^n$ by $(w/z)^{\cl n}$ in
the momentum mode contributions when the infinite sum defining
$Y(z,w)$ acts on a level-\cl state $\ket{\va}$. 
More specifically, we shall see that, when acting on such states,
this operator gives rise to poles of arbitrary order located at (see
Fig.\ \ref{fig1} below)
\[ z=w_p:=\gz^pw,\qquad 1\le p\le\cl, \]
where $\gz:=e^{2\pi i/\cl}$ and \cl denotes the eigenvalue of \K[1].
\begin{figure} \lb{fig1} \begin{center}
\setcoordinatesystem units <.3mm,.3mm> 
\unitlength.3mm
\begin{picture}(450,200)
\multiput(0,0)(250,0){2}{
         \begin{picture}(200,200) 
         \put(100,0){\vector(0,1){200}}
         \put(0,100){\vector(1,0){200}}
         \put(185,185){\line(0,1){15}}
         \put(185,185){\line(1,0){15}}
         \put(190,190){\makebox(0,0)[bl]{\large\textit{z}}}
         \end{picture}}
\put(138.50,161.95){\makebox(0,0)[t]{$w$}}

\put(104,100){\beginpicture \linethickness.8mm 
              \circulararc 360 degrees from 90 0 center at 0 0
              \circulararc 360 degrees from 60 0 center at 0 0  
              \endpicture}
\put(350,100){\beginpicture \setdashes 
              \circulararc 360 degrees from 75 0 center at 0 0
              \endpicture}

\thicklines
\put(141.50,164.95){\circle*{3}}
\put(76.60,169,82){\circle*{2}}
\put(32.33,122.11){\circle*{2}}
\put(42.03,57.75){\circle*{2}}
\put(98.40,25.21){\circle*{2}}
\put(158.98,48.99){\circle*{2}}
\put(178.16,111.18){\circle*{2}}
\put(387.50,164.95){\circle*{3}}
\put(322.60,169,82){\circle*{3}}
\put(278.33,122.11){\circle*{3}}
\put(288.03,57.75){\circle*{3}}
\put(344.40,25.21){\circle*{3}}
\put(404.98,48.99){\circle*{3}}
\put(424.16,111.18){\circle*{3}}
\put(387.50,164.95){\circle{30}}
\put(322.60,169,82){\circle{30}}
\put(278.33,122.11){\circle{30}}
\put(288.03,57.75){\circle{30}}
\put(344.40,25.21){\circle{30}}
\put(404.98,48.99){\circle{30}}
\put(424.16,111.18){\circle{30}}
\put(387.50,178.45){\vector(-1,0){5}}
\put(322.60,183,32){\vector(-1,0){5}}
\put(278.33,135.61){\vector(-1,0){5}}
\put(288.03,71.25){\vector(-1,0){5}}
\put(344.40,38.71){\vector(-1,0){5}}
\put(404.98,62.49){\vector(-1,0){5}}
\put(424.16,124.68){\vector(-1,0){5}}
\put(38.5,161.5){\vector(-1,-1){5}}
\put(59.5,140.5){\vector(1,1){5}}
\put(220,100){\makebox(0,0)[l]{$\Longrightarrow$}}
\end{picture}
\caption{Location of poles for level~$\ell=7$}
\end{center} \end{figure}
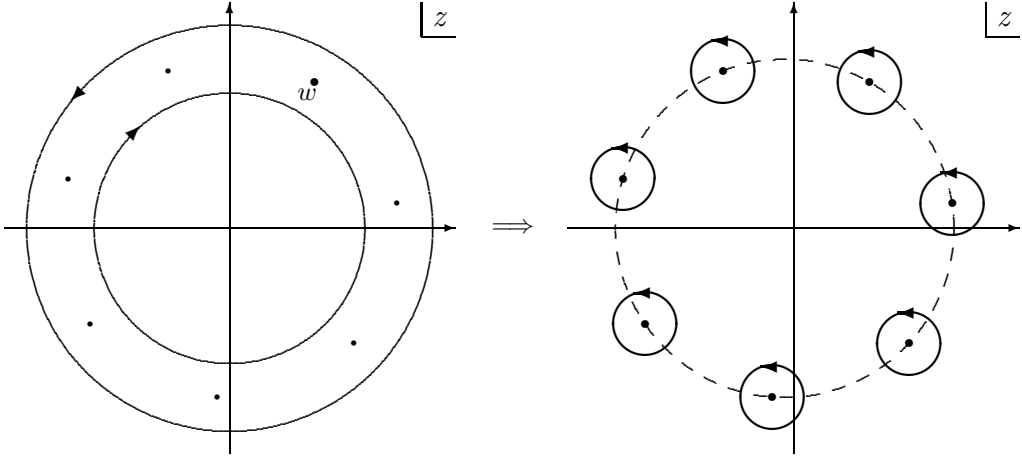
These extra poles will lead to non-local (in the sense of
quantum field theory) integrands in our final expressions.

Let us first analyze the pole that $Y(z,w)$ gives
rise to at $z=w\equiv w_\cl$; expansion around $z=w$ yields 
\[ Y(z,w)=-\frac1{(z-w)f_\cl(z,w)}, \]
where the function $f_\cl(z,w)$ does not vanish at $z=w$; explicitly,
\[ f_\cl(z,w)&=&\sum_{k\ge1}\frac1{k!}(z-w)^{k-1}
             \left(\frac{\partial^k}{\partial z^k}
                   e^{i\svd\.[\vX(z)-\vX(w)]}\right)\bigg|_{z=w}
                   \non 
           &=& \vgd\X\vP(w)
              +\frac12(z-w)[\vgd\X\vP'(w)+(\vgd\X\vP(w))^2] \non
           & &{}   +\frac16(z-w)^2[\vgd\X\vP''(w)
                                +3\vgd\X\vP(w)\vgd\X\vP'(w)
                                +(\vgd\X\vP(w))^3] 
              +\ldots, \]
with the momentum field $\vP(z)$ already defined in Eq.\ \Eq{FuVe-mom}.
When we insert the expansion of $Y(z,w)$ back into Eq.\ \Eq{adad-1},
we observe that the integrand has a pole of third order at
$z=w_\cl\equiv w$. Application of Cauchy's theorem therefore yields 
\[ \lefteqn{\sum_{\vr\in\bD}\sum_{n\in\Zn}
            \xord{\Er[1]{n}\Ema[1]{m-n}}} \hh{8mm} \non
   &=&  \sum_{\vr\in\bD}\Res[0]{w}\bigg\{
         -\frac12\frac{\partial^2}{\partial z^2}
          \left(\frac{\ord{e^{i\vr\.[\vX(z)-\vX(w)]}}}%
                     {f_\cl(z,w)}\right)\bigg|_{z=w} \non
   & &{}\hh{5mm}
        +\sum_{p=1}^{\cl-1}\Res[z=w_p]{z}\left[
          \frac{\ord{e^{i\vr\.[\vX(z)-\vX(w)]}}}{(z-w)^2}
          Y(z,w)\right]
        \bigg\}e^{-im\svd\.\vX(w)}. \lb{adad-2} \]
The first term may be further simplified by noting that the sum over
both the positive and negative roots of \bfg cancels expressions
linear in $\vr$. Hence 
\[ \sum_{\vr\in\bD}\frac{\partial^2}{\partial z^2}
         \left(\frac{\ord{e^{i\vr\.[\vX(z)-\vX(w)]}}}%
                   {f_\cl(z,w)}\right)\bigg|_{z=w}
   =\sum_{\vr\in\bD}\left[
           \frac{\ord{(\vr\X\vP(w))^2}}{\vgd\X\vP(w)}
          +\frac{\vgd\X\vP(w)}{6}  
          -\frac{\vgd\X\vP''(w)}{3(\vgd\X\vP(w))^2}    
          +\frac{(\vgd\X\vP'(w))^2}{2(\vgd\X\vP(w))^3}\right]. \lb{adad3} \]

Next recall that the physical states of a subcritical bosonic string
are finite linear combinations of states of the form
\[ \AI{i_1}{-m_1}\cdots\AI{i_M}{-m_M}
   \AI{-}{-n_1}\cdots\AI{-}{-n_N}\ket{\va}, \lb{phys-span} \] 
where \ket{\va} is any tachyonic state with \K[1]-eigenvalue \cl and the 
operators \Ai{m} and \AI{-}{m} denote the transversal and the
longitudinal DDF operators, respectively \ct{Brow72}. In
order to know the action of the Sugawara operators on arbitrary
physical states, it is therefore sufficient to work out explicitly the
action of the \sL{m}'s on a tachyonic ground state and then to
determine their commutation relations with the DDF operators.

So let us consider a state \ket{\va} satisfying $\va^2=2$ and
$\K[1]\ket{\va}=-(\vgd\X\va)\ket{\va}=\cl\ket{\va}$ for some
$\cl\in\Nn$. Evidently, \ket{\va} is a highest weight state for
$\Vir_{\sL{}}$, 
\[ \sL{m}\ket{\va}=0 \qquad \forall m>0, \nn \]
because $(\va-m\vgd)^2=2(1+\cl m)>2$ for $m>0$, but there is no physical
string state below the tachyon. For $m\ge0$, we first have to 
evaluate $Y(z,w)\ket{\va}$. We find that
\[ Y(z,w)\ket{\va}
   &=& \sum_{n\ge0}e^{in\svd\.[\vX(z)-\vX(w)]}\ket{\va} \non
   &=& \sum_{k\ge0}\sum_{n\ge0}
        \frac{n^k}{k!}\left(\frac{w}{z}\right)^{\cl n}
        \left[i\vgd\X\vX_<(z)-i\vgd\X\vX_<(w)\right]^k\ket{\va} \non
   &=& \sum_{k\ge0}
        \frac{z^{\cl(k+1)}p_k\left((w/z)^\cl\right)}%
             {k!\left[z^\cl-w^\cl\right]^{k+1}}    
        \left[i\vgd\X\vX_<(z)-i\vgd\X\vX_<(w)\right]^k\ket{\va}, \]
where
\[ i\vgd\X\vX_<(z) := \sum_{n>0}\frac1n\da{-n}z^n, \nn \]
and 
\[ p_0\equiv1,\qquad
   p_{k+1}(x) := x[(1-x)p'_k(x)+(k+1)p_k(x)]\quad\forall k\ge0. \]
The latter recursion relation follows from the formula
\[ \sum_{n\ge0}n^kx^n
   =\left(x\frac{d}{dx}\right)^k\left(\frac1{1-x}\right)
   = \frac{p_k(x)}{(1-x)^{k+1}}\qquad(|x|<1). \]
The polynomials $p_k(x)$ only have positive coefficients. Indeed, the
above recursion relations translate into
\[ p_{k+1,i}=ip_{k,i}+(k-i+2)p_{k,i-1}
   \quad\forall k>0,\ 0\le i\le k+1, \nn \]
where\footnote{By induction, it is easy to show that the coefficients
  are symmetric in the sense that $p_{k,i}=p_{k,k-i+1}\ \forall
  k>0,\ 0\le i\le k$. In particular, $p_{k,k}=p_{k,1}=1\ \forall k>0$.}
\[ p_k(x) = \sum_{i=0}^k p_{k,i}x^i. \nn \]
Hence, in particular, the polynomials cannot vanish at $x=1$ which
proves that the term 
\[ \frac{z^{\cl(k+1)}p_k\left((w/z)^\cl\right)}%
        {\left(z^\cl-w^\cl\right)^{k+1}} \nn \]
contains \cl poles at $z=w_p\equiv\gz^pw$, each of order $k+1$.

On the other hand, the expression
$[i\vgd\X\vX_<(z)-i\vgd\X\vX_<(w)]^k$ 
is a sum of terms of the form 
\[ (z^{n_1}-w^{n_1})\cdots(z^{n_k}-w^{n_k})\da{-n_1}\cdots\da{-n_k},
   \qquad n_i>0\ \forall i,\nn \]
each of them having a zero at $z=w$ of order $k$. In total,
$Y(z,w)\ket{\va}$ always has a simple pole at $z=w\equiv w_\cl$, which
was already evaluated in \Eq{adad3}, but exhibits a much more
complicated pattern at the other poles. For example, if
$(n_i,\cl)=m>0$ (highest common divisor) then the poles at $z=e^{2\pi
  ik/m}$, $1\le k\le m$, in $(z^\cl-w^\cl)^{-1}$ cancel against the
zeros in $z^{n_i}-w^{n_i}$. Up to oscillator number two, for instance,
one has the explicit formula
\[ Y(z,w)\ket{\va}
   &=& \bigg\{
        \frac{z^\cl}{z^\cl-w^\cl}
       +\frac{z^\cl w^\cl}{(z^\cl-w^\cl)^2}\left[
         (z-w)\da{-1}+\frc12(z^2-w^2)\da{-2}+\ldots\right] \non
   & &{}\hh{2mm}    
       +\frac{z^\cl w^\cl(z^\cl+w^\cl)}{2(z^\cl-w^\cl)^3}\left[
         (z-w)^2\da{-1}^2+\ldots\right]+\ldots\bigg\}\ket{\va}. \]
It is obvious from this result that a direct evaluation of
$\sL{-m}\ket{\va}$ quickly becomes unfeasible with increasing $m$.
There is, however, an elegant argument which allows us to 
shortcut this calculation and to read off the result
directly from the expression \Eq{adad-2}. We recall that
the leading oscillator contribution of a DDF operator is
\[ \Ai{-m} \sim \vgx_i\X\vga_{-m} +\ldots,\qquad
   \AI-{-m} \sim \va\X\vga_{-m} +\ldots\quad. \nn \]
Since these oscillators are linearly independent we can immediately
rewrite a given physical state in terms of DDF states simply by identifying
the leading oscillators. An important assumption here, without which this
argument would be invalid, is that there must not be any null physical state
present, because their appearance would spoil the nice oscillator
structure. Now a glance at Eq.\ \Eq{adad-2} shows that
\emph{ longitudinal oscillators are absent altogether}. This
means that the Sugawara generators when applied to any physical state
neither produce null physical states nor additional longitudinal
excitations (apart from those already contained in the initial state
\Eq{phys-span}). We conclude that the Sugawara generators can be rewritten in
terms of the transversal DDF operators alone and that the result can be
obtained by isolating those terms which do not contain $\vgd\X\vga_{-n}$
oscillators. For the second term in the Sugawara generators we
find in this way that
\[ \lefteqn{\sum_{\vr\in\bD}\sum_{n\in\Zn}
            \xord{\Er[1]{n}\Ema[1]{m-n}}} \hh{8mm} \non
   &=&  \sum_{\vr\in\bD}\Res[0]{w}\bigg\{
           \frac{w}{2\cl}\ord{(\vr\X\vP(w))^2}
          +\frac{\cl^2-1}{12\cl w} 
        +\sum_{p=1}^{\cl-1}\Res[z=w_p]{z}\left[
          \frac{\ord{e^{i\vr\.[\vX(z)-\vX(w)]}}z^\cl}%
               {(z-w)^2(z^\cl-w^\cl)}\right]\bigg\} 
          w^{\cl m} \non
        & &{} +\mbox{terms containing $\vgd\X\vga_{-n}$'s}. \lb{adad-3} \]
Note that the integrals around $z=w_p$ for the displayed terms can be
immediately performed since the integrands have only simple poles. 
The above reasoning ensures that all terms involving $\vgd$'s 
in \Eq{adad-3} must combine with the other terms precisely in such a way
that the ordinary string oscillators are replaced by DDF oscillators.
After this ``leap of faith" we arrive at
\[ \lefteqn{\sum_{\vr\in\bD}\sum_{n\in\Zn}
            \xord{\Er[1]{n}\Ema[1]{m-n}}} \hh{8mm} \non
    &=&  \sum_{\vr\in\bD}\Res[0]{w}\bigg\{
            \frac1{2\cl w}\xord{(\PI[\vr](w))^2}
            +\frac{\cl^2-1}{12\cl w} 
           +\sum_{p=1}^{\cl-1}
          \frac{w_p}{\cl(w_p-w)^2}
          \xord{e^{i\vr\.[\XI(w_p)-\XI(w)]}}\bigg\}
          w^{\cl m}    \lb{Sug-3}  \]
Our main result is thus the following new realization of the Sugawara
generators at arbitrary level:
\begin{Thm}
The operators
\[ \sL{m} 
        &=& \frac1{2(\cl+\hv)}\sum_{n\in\Zn}
              \sum_{i=1}^{d-2}\xord{\Ai[1]{n}\Ai[1]{m-n}} 
            +\frac{\hv}{2\cl(\cl+\hv)}\sum_{n\in\Zn}
              \sum_{i=1}^{d-2}\xord{\Ai{n}\Ai{\cl m-n}}
             +\frac{(\cl^2-1)(d-2)\hv}{24\cl(\cl+\hv)}\gd_{m,0} \non
        & &{}-\frac1{2\cl(\cl+\hv)}\sum_{\vr\in\bD}
              \sum_{p=1}^{\cl-1}\frac1{|\gz^p-1|^2}
              \Res[0]{w}\left\{
              w^{\cl m -1}
              \xord{e^{i\vr\.\YI{p}(w)}}\right\}, \lb{Sug-4} \]
form a Virasoro algebra with central charge 
\[ c(\cl):=\frac{\cl\dim\bfg}{\cl+\hv}. \nn \]
\end{Thm}
In deriving this result we have made use of the identity
\[ \sum_{\vr\in\bD}\xord{\PI[\vr](z)\PI[\vr](z)}
   =2\hv\sum_{i=1}^{d-2}\xord{\PI[i](z)\PI[i](z)}. \nn \]
Observe that \Eq{Sug-4} is ``doubly transcendental'' as a function 
of the ordinary string oscillators because the new coordinate field
\Eq{DDF-coo}, which itself is already a transcendental function of the
string oscillators, appears in the exponential. Moreover, this
expression is manifestly physical as it depends only on the difference
of the coordinate field. \Eq{Sug-4} contains the well known formula
\Eq{lev1-equiv} as a special case for $\ell =1$.

With the above formula, the level-\cl energy-momentum tensor
\[ \sL{}(z) := \sum_{m\in\Zn}\sL{m}z^{-\cl m}, \lb{enmom-1} \]
becomes nonlocal, to wit
\[ \sL{}(z)
        &=& \frac1{2(\cl+\hv)}
              \sum_{i=1}^{d-2}\xord{\PL[1]{i}(z^\cl)\PL[1]{i}(z^\cl)} 
            +\frac{\hv}{2\cl(\cl+\hv)}
              \sum_{i=1}^{d-2}\xord{\PI[i](z)\PI[i](z)}
             +\frac{(\cl^2-1)(d-2)\hv}{24\cl(\cl+\hv)} \non
        & &{}-\frac1{2\cl(\cl+\hv)}\sum_{\vr\in\bD}
              \sum_{p=1}^{\cl-1}\frac1{|\gz^p-1|^2}
              \xord{e^{i\vr\.[\XI(z_p)-\XI(z)]}}. \lb{enmom-2} \]

\section{General Properties and Proof of Theorem} \lb{sec:Con}
Before we prove that the expressions \Eq{Sug-4} for the level-\cl
Sugawara operators indeed satisfy the Virasoro algebra with the
correct central charge, we would like to discuss some general features
of our new formula. In the next section we will work out some explicit
examples.

Since the operators \Eq{Sug-4} are purely transversal in terms of DDF
oscillators, we immediately see that they commute with the
longitudinal DDF operators, 
\[ [\AI{-}{n},\sL{m}]=0\qquad\forall m,n\in\Zn. \]
The commutation relations with the transversal DDF operators can be
verified as follows. A straightforward calculation yields
\[ [\Aj{n},\xord{e^{i\vr\.\YI{p}(w)}}]
    =(\vr\X\vgx_j)(\gz^{pn}-1)w^n\xord{e^{i\vr\.\YI{p}(w)}}. \]
Similarly, one finds that
\[ \sum_{k\in\Zn}[\Aj{n},\xord{\Ai{k}\Ai{m-k}}]=2n\gd^{ij}\Aj{m+n}. 
    \nn \]
Inserting these results into formula \Eq{Sug-4} we obtain 
\[ [\Aj{n},\sL{m}]
        &=& \frac{n}{(\cl+\hv)}\sum_{k\in\Zn}
              \gd_{\cl k+n,0}\Aj[1]{m-k} 
            +\frac{n\hv}{\cl(\cl+\hv)}\Aj{\cl m+n} \non
        & &{}-\frac1{2\cl(\cl+\hv)}\sum_{\vr\in\bD}(\vr\X\vgx_j)
              \sum_{p=1}^{\cl-1}\frac{\gz^{pn}-1}{|\gz^p-1|^2}
              \Res[0]{w}\left\{
              w^{\cl m +n-1}
              \xord{e^{i\vr\.\YI{p}(w)}}\right\}, \lb{Sug-DDF} \]
and therefore recover the formula $[\Aj[1]{n},\sL{m}]=n\Aj[1]{m+n}$ in
agreement with Eq.\ \Eq{sem-prod}. 

It is instructive to have a closer look at the new expression for
$\sL{0}$, which reads 
\[ \sL{0}\ket\va
        &=& \bigg[\frac1{2\cl}\bva^2
             +\frac{(\cl^2-1)(d-2)\hv}{24\cl(\cl+\hv)}
             -\frac1{2\cl(\cl+\hv)}\sum_{\vr\in\bD}
              \sum_{p=1}^{\cl-1}\frac{\gz^{p\vr\.\va}}{|\gz^p-1|^2}
            \bigg]\ket\va, \lb{Sug0-1} \]
where $\bva\equiv\bvL$ denotes the projection of $\va$ (resp.\
$\vgL$) onto $\bfh^*$. Let us focus on the last term. We have the
following 
\begin{Lem} \lb{Lem1}
Let $\gz$ be a primitive \cl-th root of unity and let
$k=0,\ldots,\cl-1$. Then
\[ \sum_{p=1}^{\cl-1}\frac{\gz^{pk}}{|\gz^p-1|^2}
   =\frc1{12}(\cl^2-1)-\frc12k(\cl-k). \lb{lemma} \]
\end{Lem}
\textit{Proof:}\footnote{We would like to thank H. Samtleben for the crucial
  idea.}
With the elementary algebraic identity (for $p\not\equiv0\mod\cl$)
\[ \frac1{1-\gz^p}= - \frac1{\cl}\sum_{j=1}^{\cl-1}j\gz^{pj}, \nn \]
we immediately obtain
\[ \sum_{p=1}^{\cl-1}\frac{\gz^{pk}}{|\gz^p-1|^2}
   = \frac1{\cl^2}\sum_{i,j,p=1}^{\cl-1}ij\gz^{p(k+i-j)} 
   = \frac1{\cl^2}\sum_{i,j=1}^{\cl-1}ij
                    \sum_{p=0}^{\cl-1}\gz^{p(k+i-j)}
       -\frc14(\cl-1)^2. \nn \]
Invoking the following well-known property of sums of roots of unity,
\[ \sum_{p=0}^{\cl-1}\gz^{pn}=
   \cases{ \cl & if $n\equiv0\mod\cl$, \cr
             0 & else, } \lb{sum-zp} \]
we find that
\[ \sum_{i,j=1}^{\cl-1}ij\sum_{p=0}^{\cl-1}\gz^{p(k+i-j)}
   &=& \cl\sum_{i=1}^{\cl-1-k}i(i+k)
    +\cl\sum_{i=\cl-k}^{\cl-1}i(i+k-\cl) \non
   &=& \cl\sum_{i=1}^{\cl-1}i(i+k-\cl)
    +\cl^2\sum_{i=1}^{\cl-1-k}i \non
   &=& \cl\left[\frc16(\cl-1)\cl(2\cl-1)+\frc12(k-\cl)\cl(\cl-1)
                +\frc12\cl(\cl-1-k)(\cl-k)\right], \nn \]
and thus
\[ \sum_{p=1}^{\cl-1}\frac{\gz^{pk}}{|\gz^p-1|^2}
   &=& \frc16(\cl-1)(2\cl-1)-\frc12k(\cl-k)-\frc14(\cl-1)^2 \non
   &=& \frc1{12}(\cl^2-1)-\frc12k(\cl-k)\qquad\qquad\mbox{q.e.d.} \nn \]
By use of this result and some well-known facts about the finite
root system, we may rewrite the last term of the above
formula for $\sL{0}$ as
\[ \sum_{\vr\in\bD}\sum_{p=1}^{\cl-1}
                   \frac{\gz^{p\vr\.\va}}{|\gz^p-1|^2}
   &=& 
   \sum_{\vr\in\bD_+}\left[\frc16(\cl^2-1)
                           -(\vr\X\va)(\cl-(\vr\X\va))\right] \non 
   &=& \frac{(\cl^2-1)(d-2)\hv}{12}-2\cl\bvr\X\bva+\hv\bva^2, \nn \]
where $\bvr$ denotes the Weyl vector for the finite
subalgebra.\footnote{Note that the term linear in $k=\vr\X\va$ does
  not drop out upon summation over the roots but rather reproduces the
  Weyl vector. This is due to the fact that the Lemma is valid only
  for $0\le k\le\cl-1$ and thus different values of $\vr\X\va$ have to
  be transported into this range by multiples of \cl.} 
If we insert this into \Eq{Sug0-1} we arrive at the formula
\[ \sL{0}\ket\va
    =\frac{(\bva+2\bvr)\X\bva}{2(\cl+\hv)}\ket\va, \lb{Sug0-2} \]
in agreement with \cite[Lemma12.8.b)]{Kac90}. Note that we have not
employed any properties of the affine Casimir operator in our
calculation. 

It remains to verify that the operators \Eq{Sug-4} really
do satisfy the Virasoro algebra with the correct central charge. 
To this aim we split the Sugawara operators and introduce the 
following operators: 
\[ \tsL{m} = \LV1{m} + \LV2{m} + \LV3{m}, \lb{Sug-5} \]
with
\[ \LV1{m} 
    &:=& \frac1{2\cl}\sum_{n\in\Zn}
         \sum_{i=1}^{d-2}\xord{\Ai[1]{n}\Ai[1]{m-n}}, \non
   \LV2{m} 
    &:=& \frac{\hv}{2\cl(\cl+\hv)}
         \sum_{n\in\Zn \atop n\not\equiv0(\cl)}
         \sum_{i=1}^{d-2}\xord{\Ai{n}\Ai{\cl m-n}}, \non
   \LV3{m} 
    &:=&{}-\frac1{2\cl(\cl+\hv)}\sum_{\vr\in\bD}
           \sum_{p=1}^{\cl-1}\frac1{|\gz^p-1|^2}
           \Res[0]{w}\left\{  w^{\cl m -1}
           \xord{e^{i\vr\.\YI{p}(w)}}\right\}. \lb{Sug-6} \]
Observe that we have absorbed all terms involving \Ai[1]{n} into
\LV1{m} so that the prefactor is ``renormalized" to $(2\cl)^{-1}$
with respect to \Eq{Sug-4}, because these operators commute
with the DDF oscillators \Ai{n} with $n\not\equiv0(\cl)$.
We obviously have (with $\K[1]=\cl$)
\[ [\LV1{m},\LV1{n}] 
    &=& (m-n)\LV1{m+n}+\frac{d-2}{12}(m^3-m)\gd_{m+n,0}, \non {}
   [\LV1{m},\LV2{n}] 
    &=& [\LV1{m},\LV3{n}]=0.  \lb{Sug-Vir1} \]
It is equally straightforward to show that
\[ [\LV2{m},\LV2{n}] 
    = (m-n)\frac{\hv}{\cl+\hv}\LV2{m+n} 
      +\frac{(d-2)(\cl-1)(\hv)^2}{(\cl+\hv)^2} 
       \left(m^3+\frac{m}{\cl}\right)\gd_{m+n,0}  \lb{Sug-Vir2}  \]
and 
\[ [\LV2{m},\LV3{n}] 
    = (m-n)\frac{\hv}{\cl+\hv}\LV3{m+n}. \lb{Sug-Vir3} \]
The remaining commutator requires more work. For its evaluation we
need the operator product 
\[ \xord{e^{i\vr\.\YI{p}{(z)}}} \xord{e^{i\vs\.\YI{q}{(w)}}}=
  \left[\frac{(z_p-w_q)(z-w)}{(z_p-w)(z-w_q)}\right]^{\vr\.\vs}
  \xord{e^{i\vr\.\YI{p}{(z)} + i\vs\.\YI{q}{(w)}}}. \lb{op-prod} \] 
We write
\[ [\LV3{m},\LV3{n}]
    = \sum_{\vr,\vs\in\bD}\sum_{p,q=1}^{\cl-1}I(\vr,\vs,p,q), \nn \]
with
\[ I(\vr,\vs,p,q)
    &:=&\frac1{4\cl^2(\cl+\hv)^2}
        \frac1{|\gz^p-1|^2|\gz^q-1|^2}\times \non
    & &\times
        \Res[0]{w}\sum_{a=1}^\cl\Res[w_a]{z}
        \left\{w^{\cl n -1}z^{\cl m -1}  
               \left[\frac{(z_p-w_q)(z-w)}%
                          {(z_p-w)(z-w_q)}\right]^{\vr\.\vs}
               \xord{e^{i\vr\.\YI{p}(z)+i\vs\.\YI{q}(w)}}\right\}. \nn \]
Inspection of the term in curly brackets reveals poles of order 1,2
and 4, depending on the values of $\vr\X\vs$ and $p,q \in \{1,\ldots,\cl
-1\}$. As usual, there is no contribution for $\vr\X\vs =0$. Another
useful observation is that 
\[ I(\vr,\vs,p,q)=I(\vr,-\vs,p,\cl-q)\qquad\forall \vr,\vs,p,q, \]
It is therefore sufficient to consider the cases
$\vr\X\vs=-1$ ($\Leftrightarrow\vr+\vs\in\bD$) and 
$\vr\X\vs=-2$ ($\Leftrightarrow\vs=-\vr$); these lead to the
following results: 
\ben
\item If $\vr\X\vs=-1$ and $p=q$ (pole of order 2), then
\[ \sum_{\vr\in\bD}\sum_{p=1}^{\cl-1}I(\vr,\vs,p,p)
    =(n-m)\frac1{4(\cl+\hv)}\LV3{m+n}, \nn \]
after partial integration. For simply laced algebras and given
$\vr\in\bD$, there are always $2\hv-4$ roots $\vs$ such that
$\vr\X\vs=-1$ (or $+1$)\footnote{By Weyl invariance it is sufficient
  to prove the statement for the highest root $\vgt$. From the
  definition of the Coxeter number and the Weyl vector we have
  $$ 2(\hv-1) = 2\bvr\X\vgt = \sum_{\vs\in\bD_+} \vs\X\vgt. $$
  The only contributions in the sum arise from the terms with
  $\vs\X\vgt=1$ (whose number we wish to compute) and with
  $\vs\X\vgt=2$. However, the only positive root for which
  $\vs\X\vgt=2$ is $\vs=\vgt$, whence the result.}. Hence
\[ \sum_{\vr,\vs\in\bD \atop \vr\.\vs=-1}
   \sum_{p=1}^{\cl-1}I(\vr,\vs,p,p)
    =(m-n)\frac{2-\hv}{2(\cl+\hv)}\LV3{m+n}, \nn \]
\item If $\vr\X\vs=-1$ and $p\neq q$ (poles of order 1), then
\[ I(\vr,\vs,p,q)=-I(\vs,\vr,q,p), \nn \]
which vanishes upon (symmetric) summation over $\vr,\vs,p,q$.
\item If $\vr\X\vs=-2$ and $p=q$ (pole of order 4), then
\[ \sum_{\vr\in\bD}\sum_{p=1}^{\cl-1}I(\vr,-\vr,p,p)
    =(m-n)\frac{\cl}{2(\cl+\hv)}\LV2{m+n}
      +\frac{(d-2)\cl(\cl-1)\hv}{24(\cl+\hv)^2} 
       \left(m^3+\frac{m}{\cl}\right)\gd_{m+n,0}, \nn \]
after partial integration and use of Lemma \ref{Lem1}.
\item If $\vr\X\vs=-2$ and $p\neq q$ (poles of order 2), then
\[ \sum_{\vr\in\bD}\sum_{p,q=1 \atop p\neq q}^{\cl-1}I(\vr,-\vr,p,q)
    =(m-n)\frac{\cl-2}{2(\cl+\hv)}\LV3{m+n}, \nn \]
after partial integration.
\een
Hence 
\[ [\LV3{m},\LV3{n} ] &=& (m-n) \frac{\cl}{\cl +\hv} \LV2{m+n}
    + (m-n) \frac{\cl-\hv}{\cl + \hv} \LV3{m+n} \non
  && + \frac{(d-2)\cl (\cl -1)\hv}{12 (\cl + \hv )^2} 
   \left( m^3 + \frac{m}{\cl} \right) \gd_{m+n,0}. \lb{Sug-Vir4} \]
Adding up all contributions we get 
\[ [\tsL{m},\tsL{n}]=(m-n)\tsL{m+n}
                     +\left(\frac{c}{12}m^3+bm\right)\gd_{m,0}, \nn \]
with 
\[ b:=-\frac{d-2}{12}+\frac{(d-2)(\cl-1)\hv}{12\cl(\cl+\hv)} 
    =-\frac{(d-2)(\cl^2+\hv)}{12\cl(\cl+\hv)}; \nn \]
the central charge $c$ given by
\[ c&=&d-2+\frac{(d-2)(\cl-1)(\hv)^2}{(\cl+\hv)^2}
          +\frac{(d-2)\cl(\cl-1)\hv}{(\cl+\hv)^2} \non
    &=&\frac{(d-2)\cl(1+\hv)}{\cl+\hv}=\frac{\cl\dim\bfg}{\cl+\hv}, \nn
    \]
in agreement with \Eq{Sug-cc}. Since this Virasoro algebra has not yet
the standard form, we have to shift $\tsL{0}$. Doing this we arrive at the
desired result,
\[ \sL{m}&:=&\tsL{m}+\frac{c+12b}{24}\gd_{m,0} \non
         &=&\tsL{m}+\frac{(\cl^2-1)(d-2)\hv}{24\cl(\cl+\hv)}\gd_{m,0}.
   \nn \]

Finally we would like to mention that expressions analogous to
\Eq{Sug-4} in terms of DDF oscillators also exist for 
the step operators $\Er[1]{m}$. They read 
\[ \Er{m}=\Res[0]{z}\left\{
              \xord{z^{m+1+\vr\.\vp}
                    e^{i[\vr-(\vr\.\vp+1)\vkl]\.\XI(z)}}\right\} 
   \lb{DDF-Er}. \]
One can show that these operators indeed satisfy the commutation
relations \Eq{CWcom-2} and \Eq{CWcom-3}. In addition, they allow a
direct verification of the semidirect product \Eq{sem-prod}. 

\section{Examples} \lb{sec:Sum}
In this section we present some examples. As already mentioned
we will restrict attention to the Lie algebras \8, \9 and \0.

When expanding the exponential operator in the new formula \Eq{Sug-4}, 
we notice that $\sL{n}$ involves linear combinations of the form
\[ \sum_{m_1,\ldots,m_M\not\equiv0(\cl) \atop m_1+\ldots+m_M=\cl n}
   \frac1{m_1\cdots m_M}
   \sT_{j_1\ldots j_M}(\va;\cs{m_1},\ldots,\cs{m_M})
   \AI{j_1}{-m_1}\cdots\AI{j_M}{-m_M}, \lb{wsum-1} \]
with
\[ \sT_{j_1\ldots j_M}(\va;\cs{m_1},\ldots,\cs{m_M}) :=
    \sum_{p=1}^{\cl-1}\sum_{\vr\in\bD}
    \frac{\gz^{p\vr\.\va}}{|\gz^p -1|^2}
    (\gz^{pm_1}-1)\cdots(\gz^{pm_M}-1) \, r^{j_1}\cdots r^{j_M};
    \lb{wsum-2} \]
here $\cs{m}$ is a coset representative for $m$, i.e., $m=\cs{m}+k\cl$
for some $k\in\Zn$, $\cs{m}\in\{1,\ldots,\cl-1\}$, and $r^j$ denotes the
$j$-th component of the root $\vr$ with respect to some basis $\{\ve_j|1\le
j\le d-2\}$ of $\bfh^*$. 
Since $M\ge2$, the tensors can be simplified by writing
\[  N(\vr\X\va;\cs{m_1},\ldots,\cs{m_M}) &:=&
    \sum_{p=1}^{\cl-1}
    \frac{\gz^{p\vr\.\va}}{|\gz^p -1|^2}
    (\gz^{pm_1}-1)\cdots(\gz^{pm_M}-1) \non
    &=&
    -\sum_{p=1}^{\cl-1}
    \sum_{k_1=0}^{\cs{m_1}-1}\sum_{k_2=0}^{\cs{m_2}-1}
    \gz^{p(\vr\.\va+k_1+k_2+1)}
    (\gz^{pm_3}-1)\cdots(\gz^{pm_M}-1). \lb{wsum-3} \]
Invoking \Eq{sum-zp} we conclude that the numbers
$N(\va\X\vr;\cs{m_1},\ldots,\cs{m_M})$ are always real integers. 
Hence further evaluation of the tensors $\sT_{j_1\ldots j_M}$
necessitates the computation of weighted sums over tensor products of
real roots of the following type: 
\[ \sum_{\vr\in\bD}N(\vr\X\va)\vr\XO\ldots\XO\vr, \lb{wsum-4} \]
with $N(\vr\X\va)\in\Zn$. Such sums have not been considered in the 
literature so far, except in the simplest situation where
$N=\mathrm{const}$. In this case the sums become invariant tensors w.r.t.\ 
the full Weyl group of the finite Lie algebra under consideration.
E.g., for $\8$ we have the following formula for the unweighted sums over
tensor products up to six tensor factors:
\[ \sum_{\vr\in\bD} r^{j_1}\cdots r^{j_{2k}}=
   2^{4-k} \big( 7+2^{2k-3} \big) \gd_{(j_1 j_2} \cdots 
    \gd_{j_{2k-1} j_{2k})} \FOR{ $k$=1, 2, 3},    \lb{uwsum} \]
where ${}_{(\ldots)}$ denotes symmetrization with strength one
and tensors with an odd number of indices vanish (this is no longer
true for the weighted sums \Eq{wsum-4}).
The simple result \Eq{uwsum} is explained by the absence of invariant
tensors other than $\gd_{ij}$ for $k\leq 3$ for the Weyl group 
$\fW(\8)=D_4(2) \XO \Zn_2^2$. For $k\geq 4$, new invariants
appear in accordance with the general theory since the exponents
of \8 are 1, 7, 11, 13, 17, 19, 23, 29 \ct{Slod96}. 
 
It is clear that the presence of the factor $N(\vr\X\va)$ in
$\sT(\va)$ breaks the symmetry under the full affine Weyl group down
to that subgroup which preserves $\va$; this is just the (finite)
little Weyl group $\fW(\va,\vgd)$ introduced in \ct{GebNic95}.
As a consequence, the results will be the more cumbersome the
smaller $\fW(\va,\vgd)$ becomes. For the examples to be presented 
below we have therefore evaluated the relevant sums \Eq{wsum-4} on the 
computer. Inspection of the explicit examples suggests that 
it may be difficult to find closed (or at least more elegant) 
general expressions for them, and we have not tried to do so.

Let us illustrate the new formula and the above remarks with some
examples for the exceptional Lie algebra $\bfg = \8$ with affine
extension $\fg=\9$ and hyperbolic extension $\hfg=\0$. In this
(unique) case the finite root lattice is selfdual, and consequently
the extended affine root lattice coincides with the weight lattice of \0
which is just the unique even selfdual Lorentzian lattice $\II$.
As a partial check on our results, we have have recalculated
(and reobtained) \Eq{example-1} below directly by means of formula (4.60)
in \ct{GebNic95} (the Lie algebra analog of \Eq{Sug-def}), i.e.,
without use of \Eq{Sug-4}. Doing the calculation in this ``old" way 
is impossible without massive use of algebraic computer programs, 
whereas the new formula requires substantially less effort.
In fact, knowing the tensors \Eq{wsum-2} (for this we must still rely 
on the computer), the calculation can be done by hand. 

As our first example, we choose the fundamental dominant weight
$\vgL_1=2\vr_{-1}+\vr_0+3\vgd$ of level~$\cl=2$ with associated
tachyonic vector $\va_1=\vgL_1-2\vgd$. We identify the polarization
vectors $\vgx_i$ with the orthonormal basis vectors $\ve_i$.  The
little Weyl group is $\fW(\vgL_1,\vgd) = \fW (E_7)\XO \Zn_2 = %
C_3(2)\XO\Zn_2^2$.  
An exceptional property of the level-2 sector is the vanishing of all
tensors with an odd number of indices; this feature will be lost for
higher levels $|\cl|>2$. With the notation $A_{-m}^i\equiv\Ai[2]{-m}$
we find
\[
\sL[2]{-1}\ket{\va_1} &=& \bigg\{
  \frc{3}{16}\sum_{i=1}^7A_{-1}^iA_{-1}^i 
 +\frc{7}{16}A_{-1}^8A_{-1}^8 
 +\frc12\sqrt{2}A_{-2}^8 \bigg\}\ket{\va_1}, \non
\sL[2]{-1}\sL[2]{-1}\ket{\va_1} &=& \bigg\{
  \frc{11}{64}\sum_{i=1}^7A_{-3}^iA_{-1}^i
 +\frc{11}{256}\sum_{i,k=1}^7(A_{-1}^i)^2(A_{-1}^k)^2
 +\frc{3}{16}\sqrt{2}\sum_{i=1}^7A_{-2}^8(A_{-1}^i)^2 \non
&&\quad 
 +\frc{15}{128}\sum_{i=1}^7(A_{-1}^8)^2(A_{-1}^i)^2 
 +\frc{35}{64}A_{-3}^8A_{-1}^8
 +\frc{1}{2}(A_{-2}^8)^2
 +\frc{7}{16}\sqrt{2}A_{-2}^8(A_{-1}^8)^2\non
&&\quad 
 +\frc{35}{256}(A_{-1}^8)^4 
 +\frc12\sqrt{2}A_{-4}^8\bigg\}\ket{\va_1}, \non
\sL[2]{-2}\ket{\va_1} &=& \bigg\{
 \frc{7}{16}\sum_{i=1}^7A_{-3}^iA_{-1}^i
 +\frc{1}{4}\sum_{i=1}^8A_{-2}^iA_{-2}^i 
 -\frc{1}{64}\sum_{i,k=1}^7(A_{-1}^i)^2(A_{-1}^k)^2\non
&&\quad
 +\frc{3}{32}\sum_{i=1}^7(A_{-1}^8)^2(A_{-1}^i)^2
 +\frc{29}{48}A_{-3}^8A_{-1}^8
 +\frc{5}{192}(A_{-1}^8)^4
 +\frc12\sqrt{2}A_{-4}^8 \bigg\}\ket{\va_1}, \non
\sL[2]{-3}\ket{\va_1} &=& \bigg\{
  \frc{9}{20}\sum_{i=1}^7A_{-5}^iA_{-1}^i 
 +\frc{1}{2}\sum_{i=1}^8A_{-4}^iA_{-2}^i
 +\frc{11}{48}\sum_{i=1}^7A_{-3}^iA_{-3}^i
 -\frc{1}{48}\sum_{i,k=1}^7A_{-3}^iA_{-1}^i(A_{-1}^k)^2\non
&&\quad
 +\frc{1}{16}\sum_{k=1}^7  A_{-3}^8A_{-1}^8(A_{-1}^k)^2
 +\frc{1}{16}\sum_{i=1}^7  A_{-3}^iA_{-1}^i(A_{-1}^8)^2
 -\frc{1}{64}\sum_{i,k=1}^6(A_{-1}^i)^2(A_{-1}^k)^2(A_{-1}^7)^2\non
&&\quad
 +\frc{1}{64}\sum_{i,k=1}^6(A_{-1}^i)^2(A_{-1}^k)^2(A_{-1}^8)^2
 +\frc{1}{32}\sum_{i=1}^6(A_{-1}^i)^2(A_{-1}^7)^2(A_{-1}^8)^2
 -\frc{1}{96}\sum_{i,k=1}^6(A_{-1}^i)^2(A_{-1}^k)^4\non
&&\quad
 +\frc{1}{48}\sum_{i=1}^6(A_{-1}^i)^4(A_{-1}^7)^2
 -\frc{1}{192}\sum_{i=1}^6(A_{-1}^i)^2(A_{-1}^7)^4
 +\frc{1}{64}\sum_{i=1}^6(A_{-1}^i)^2(A_{-1}^8)^4\non
&&\quad
 +\frc{1}{64}(A_{-1}^7)^2(A_{-1}^8)^4
 +\frc{1}{64}(A_{-1}^8)^2(A_{-1}^7)^4
 +\frc{1}{120}\sum_{i=1}^6(A_{-1}^i)^6
 +\frc{11}{20}A_{-5}^8A_{-1}^8
 +\frc{37}{144}A_{-3}^8A_{-3}^8\non
&&\quad
 +\frc{5}{144}A_{-3}^8(A_{-1}^8)^3
 -\frc{1}{320}(A_{-1}^7)^6
 -\frc{1}{2880}(A_{-1}^8)^6
 -\frc{1}{4}\prod_{i=1}^6A_{-1}^i
 +\frc12\sqrt{2}A_{-6}^8 \bigg\}\ket{\va_1}. \lb{example-1} \]
Our second example is the fundamental dominant weight $\vgL_8$ of level
$\cl=3$ with associated tachyonic vector $\va_8=\vgL_8-2\vgd$
and little Weyl group $\fW(\vgL_8,\vgd)= \fW(A_8)= S_8$. 
In terms of our standard basis of orthonormal polarization 
vectors we found the results 
\[
\sL[3]{-1}\ket{\va_8} &=& \bigg\{
  \frc{1}{6}\sum_{i=1}^8A_{-3}^i 
 +\frc{7}{22}\sum_{i=1}^8A_{-2}^iA_{-1}^i 
 -\frc{1}{264}\sum_{i,j,k=1}^8(1-6\gd_{ij}+12\gd_{ij}\gd_{jk}\gd_{ki})
 A_{-1}^iA_{-1}^jA_{-1}^k 
 \bigg\}\ket{\va_8}, \non 
\sL[3]{-2}\ket{\va_8} &=& \bigg\{
  \frc{1}{6}\sum_{i=1}^8A_{-6}^i 
 +\frc{17}{55}\sum_{i=1}^8A_{-5}^iA_{-1}^i 
 +\frc{27}{88}\sum_{i=1}^8A_{-4}^iA_{-2}^i 
 +\frc{1}{6}\sum_{i=1}^8(A_{-3}^i)^2 \non
&&\phantom{\bigg\{}
 -\frc{1}{352}\sum_{i,j,k=1}^8(1-4\gd_{ij}-2\gd_{jk}+12\gd_{ij}\gd_{jk})
  A_{-4}^iA_{-1}^jA_{-1}^k \non
&&\phantom{\bigg\{}
 +\frc{1}{2112}\sum_{i,j,k=1}^8(1-6\gd_{ij}+12\gd_{ij}\gd_{jk})
  A_{-2}^iA_{-2}^jA_{-2}^k \non
&&\phantom{\bigg\{}
 -\frc{1}{704}\sum_{i,j,k,l=1}^8(\gd_{ij}+\gd_{kl}+4\gd_{jk}-12\gd_{ij}\gd_{jk}
  -12\gd_{jk}\gd_{kl} \non
&&\phantom{\bigg\{\qquad}
+24\gd_{ij}\gd_{jk}\gd_{kl}-4\gd_{ij}\gd_{kl}
  -8\gd_{ik}\gd_{jl})
  A_{-2}^iA_{-2}^jA_{-1}^kA_{-1}^l \non
&&\phantom{\bigg\{}
 +\frc{1}{2816}\sum_{i,j,k,l,m=1}^8(1-8\gd_{ij}-12\gd_{jk}+24\gd_{ij}\gd_{jk}
  +16\gd_{jk}\gd_{kl}-32\gd_{ij}\gd_{jk}\gd_{kl}-8\gd_{jk}\gd_{kl}\gd_{lm}
  \non
&&\phantom{\bigg\{\qquad}
-16\gd_{ij}\gd_{jk}\gd_{kl}\gd_{lm}
+64\gd_{ij}\gd_{kl}+16\gd_{jk}\gd_{lm}
  +72\gd_{ij}\gd_{jk}\gd_{lm}+48\gd_{ij}\gd_{kl}\gd_{lm})
  A_{-2}^iA_{-1}^jA_{-1}^kA_{-1}^lA_{-1}^m \non
&&\phantom{\bigg\{}
 -\frc{1}{42240}\sum_{i,j,k,l,m,n=1}^8(1-15\gd_{ij}+40\gd_{ij}\gd_{jk}
  -60\gd_{ij}\gd_{jk}\gd_{kl}+144\gd_{ij}\gd_{jk}\gd_{kl}\gd_{lm}
  -144\gd_{ij}\gd_{jk}\gd_{kl}\gd_{lm}\gd_{mn}\non
&&\phantom{\bigg\{\qquad}
  +80\gd_{ij}\gd_{kl}\gd_{mn}
  +320\gd_{ij}\gd_{jk}\gd_{lm}\gd_{mn})
  A_{-1}^iA_{-1}^jA_{-1}^kA_{-1}^lA_{-1}^mA_{-1}^n 
 \bigg\}\ket{\va_8}, \lb{example-2} \]  
where now $A_{-m}^i\equiv\Ai[3]{-m}$. This  
example illustrates that terms with an odd number of DDF oscillators  
need not vanish in general. The invariance of under the little Weyl
group $S_8$ can be made manifest by switching to a non-orthonormal and 
$S_8$ invariant basis of polarization vectors. This leads to slightly
simpler expressions. However, this simplicity is an artefact caused by
the size of the little Weyl group, which generically becomes trivial. 
In addition we note that, by \Eq{wsum-1}, even the operator $\sL{-1}$
will involve oscillator number $m_1+\ldots+m_M = \cl$ and thus an
exponentially growing number of terms with increasing level~\cl.

\section{Outlook}
The above results nicely display the increasing ``anisotropy"
of the root space representations in terms of the DDF basis
with increasing level, a feature which we have already stressed before
and which can be traced to the decrease (and eventual triviality) of
the little Weyl group at higher level. While the further exploration 
of higher-level root spaces by direct methods as in \ct{GebNic95}
seems prohibitively difficult, prospects are much brighter with our
new formula \Eq{Sug-4}. What is still missing at this point is an
analogous and similarly explicit expression for the full coset generators
\[ \sK{m} := (\sL[1]{m}\XO\1\XO\cdots\XO\1) + \ldots +
   (\1\XO\cdots\XO\1\XO\sL[1]{m})-\sL{m}.  \lb{coset}   \] 
While the last term involves transversal DDF operators only
by \Eq{Sug-4}, we have checked that the other terms will 
introduce a dependence on the longitudinal DDF operators.
Since the operators $\sK{m}$ commute with the affine subalgebra
this might also shed some light on the long-standing problem
of finding explicit expressions for the (non-polynomial) higher
order Casimir invariants of affine algebras. We hope to come back
soon to these issues in another publication.

\vspace*{5ex} \noi
\textbf{Acknowledgments:} We are very grateful to P.~Slodowy for sharing 
with us his expertise on Weyl groups and their invariant tensors.
We would also like to thank J.~Fuchs for discussions about 
the representation theory of affine Lie algebras.

\end{document}